\documentclass{ws-procs9x6}
\input{psfig}
\def\approxgt{\mathrel{\hbox{\rlap{\lower.55ex \hbox {$\sim$}}
        \kern-.3em \raise.4ex \hbox{$>$}}}}
\def\approxlt{\mathrel{\hbox{\rlap{\lower.55ex \hbox {$\sim$}}
        \kern-.3em \raise.4ex \hbox{$<$}}}}

\begin{document}

\title{Transmission- to reflection-dominated transition in Seyfert~2
galaxies: the XMM-Newton view\footnote{\uppercase{T}his work is based on
observations obtained with \uppercase{XMM}-Newton, and
\uppercase{ESA}
science mission with instruments and contributions directly funded by
\uppercase{ESA}
Member States and the \uppercase{USA} (\uppercase{NASA}).}}

\author{M.~Guainazzi, P.~Rodriguez-Pascal}

\address{XMM-Newton Science Operations Center, RSSD of ESA, VILSPA, Apartado
50727,
E-28080 Madrid, Spain, E-mail: {\tt mguainaz@xmm.vilspa.esa.es}}

\author{A.C.~Fabian, K.~Iwasawa}

\address{Institute of Astronomy, Madingley Road, Cambridge, CB3 0HA}

\author{G.Matt}

\address{Dipartimento di Fisica ``E.Amaldi", Universit\`a ``Roma Tre",
Via della Vasca Navale, I-00184 Roma, Italy}

\author{F.Fiore}

\address{INAF-Osservatorio Astronomico di Roma, via Frascati 33,
Monteporzio-Catone (RM), 00040, Italy}


\maketitle

\abstracts{
In this paper we present
the current status of a XMM-Newton
program to observe an optically-defined, complete and unbiased
sample of Compton-thick Seyfert~2 galaxies.
The main goal of this project is the
measurement of the occurrence rate of transition between
transmission- ({\it i.e.} Compton-thin), and reflection-dominated
spectral states. These transitions potentially provide information
on the distribution of the obscuring matter surrounding the nucleus, and on the
duty-cycle
of the AGN activity. With about 2/3 of the whole sample being
observed, we detected 1 further transition out of
8 observed objects, confirming previous suggestions
that these transitions occur on time-scales $\sim$50--100 years.
}

\section{The unstable temper of Compton-thick Seyfert 2s}

Our group is pursuing a XMM-Newton survey of
an optically defined,
complete sample of 
X-ray obscured nearby Seyfert galaxies,
classified as Compton-thick according to
observations prior to the launch of
{\it Chandra} and XMM-Newton. The
main goal of this project is to measure the occurrence rate of
transitions between ``transmission-'' and ``reflection-dominated''
spectral states (Matt et al. 2003, and references therein).
Traditional
wisdom associates ``reflection-dominated'' states to
AGN covered by a Compton-thick
({\it i.e.} $N_H \ge \sigma_t^{-1} \simeq 1.5 \times 10^{24}$~cm$^{-2}$)
absorber,
which totally suppresses the AGN emission below 10~keV.

The discovery of ``transmission-'' to ``reflection-dominated''
state transitions (simply {\it transitions} hereafter:
Guainazzi et al. 2002; Guainazzi 2002)
fundamentally challenges this traditional wisdom, as well as -
in a broader astrophysical context - a ``static
interpretation'' of Seyfert Unification Models.
In at least two out of the five
known cases (Guainazzi 2002, Gilli et al. 2000),
these transitions are best explained with variations
of the AGN intrinsic power by at least one order
of magnitude on timescales $\simeq$years, as in
the Narrow Line Seyfert~1 Galaxy NGC~4051
(Uttley et al. 1999). At the same time, these
results are consistent with the
extension of the
standard Unification Models proposed by
Matt (2000), whereby the nuclear,
pc-scale torus is responsible for the Compton-thick
absorption only, whereas the Compton-thin matter
is located at much larger distances, possibly associated
to the host galaxy rather then to the nuclear
environment.

In this paper we present preliminary results of this
study, with 8 out of the foreseen 12 targets being
already observed.

\section{Th XMM-Newton results}

\begin{figure}[ht]
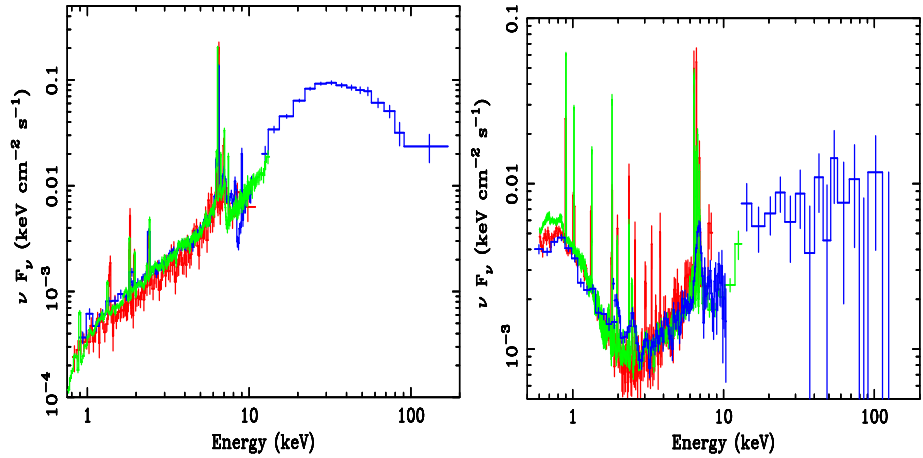

\hbox{
\psfig{figure=fig1a.ps,height=6.0cm,width=6.0cm,angle=-90}
\psfig{figure=fig1b.ps,height=6.0cm,width=6.0cm,angle=-90}
}
\caption{X-ray Spectral Energy Distributions of
the Circinus Galaxy ({\it left}) and
NGC~1068 ({\it right}) as derived from
their ASCA, BeppoSAX and XMM-Newton observations \label{fig1}}
\end{figure}

These transitions are not ubiquitous on the time scale
roughly defined by the average interval between the
ASCA/BeppoSAX, and the XMM-Newton/{\it Chandra} observations
(a few years). The Circinus Galaxy and NGC~1068, the closest
and best studied Compton-thick AGN known, exhibit a consistently
reflection-dominated spectrum across their whole recorded
X-ray history (Fig.\ref{fig1}), with very little spectral
changes.

\begin{figure}[ht]
\hbox{
\psfig{figure=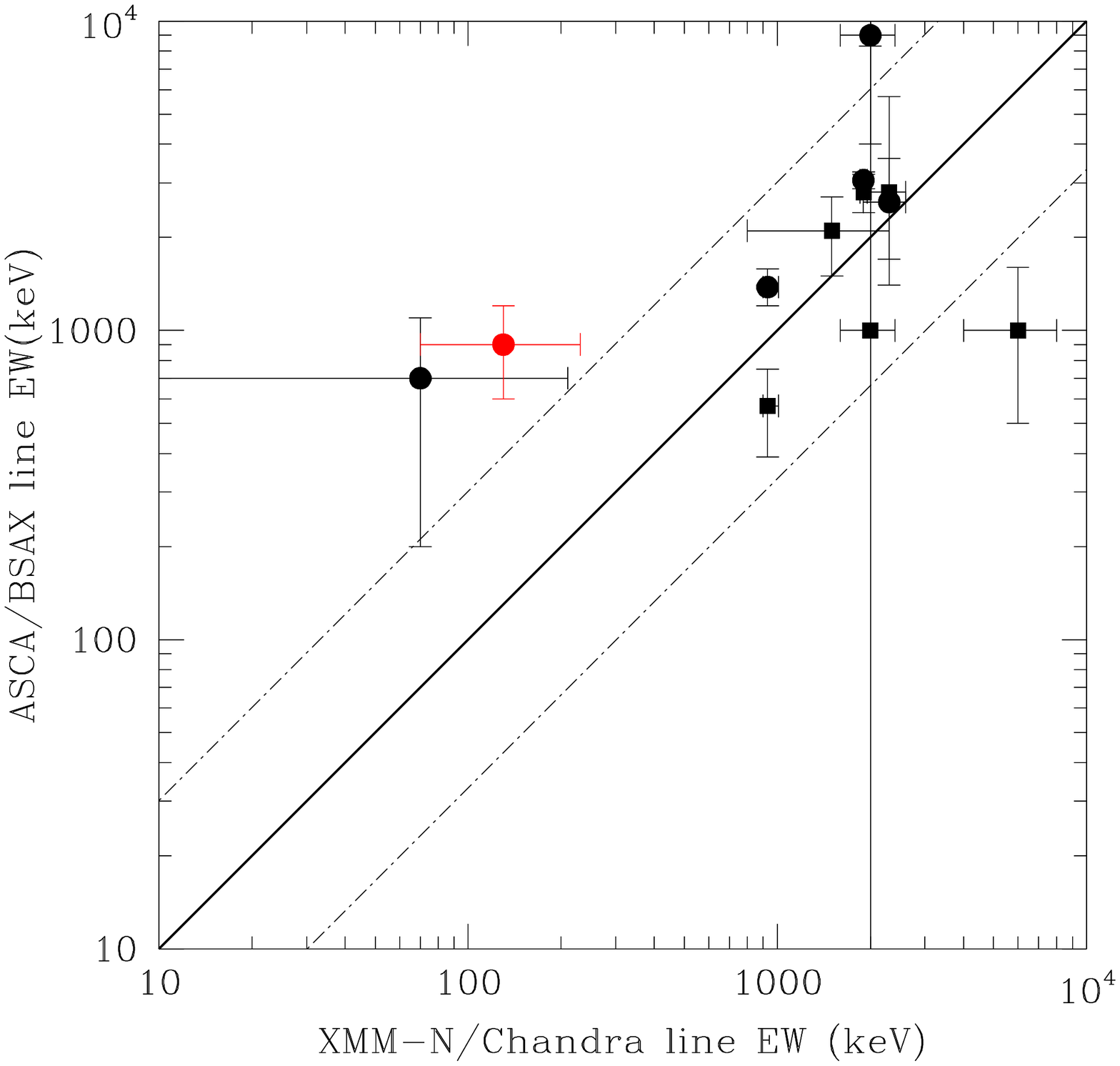,height=6.0cm,width=6.0cm}
\psfig{figure=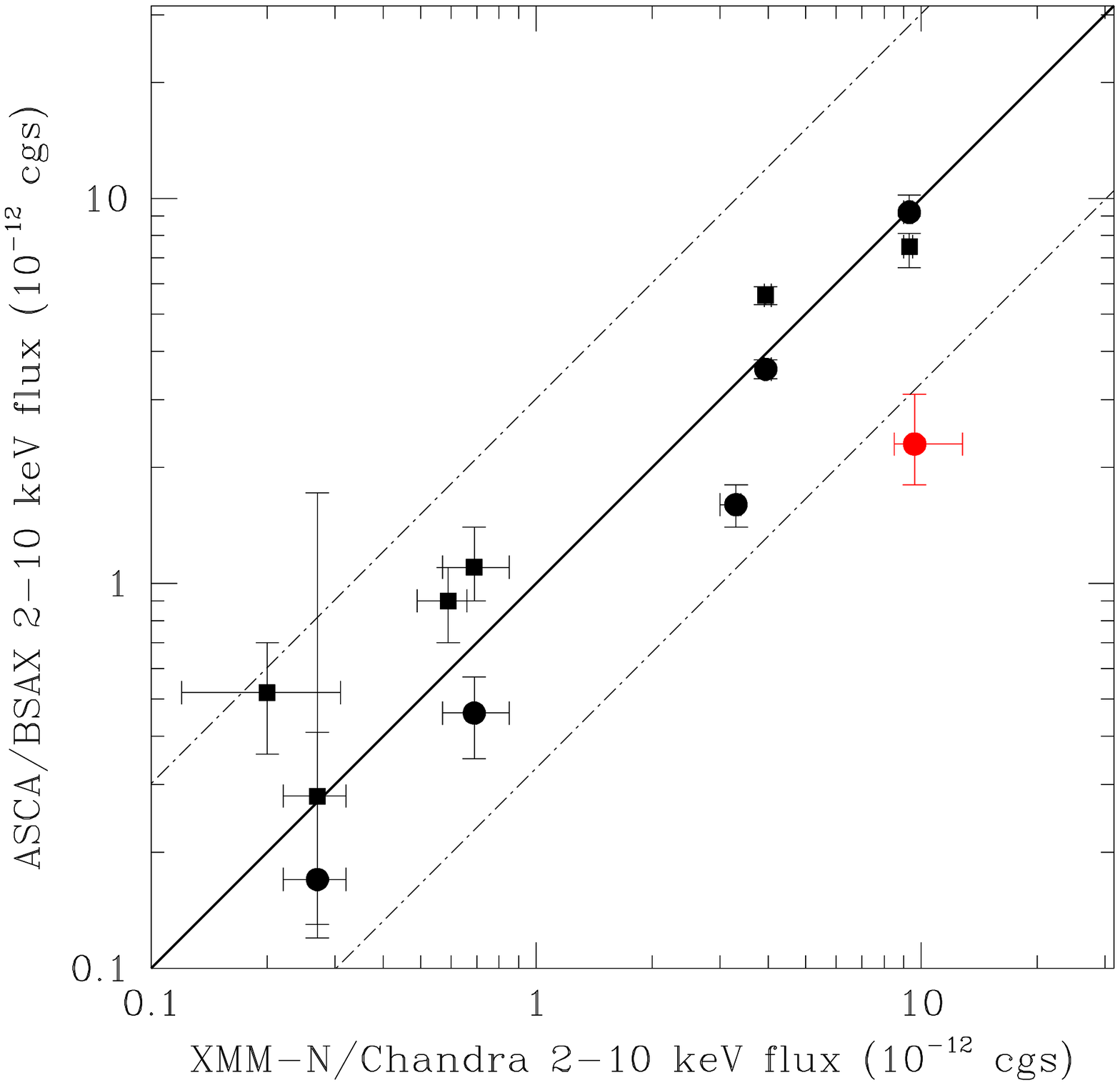,height=6.0cm,width=6.0cm}
}
\caption{Comparison between measurements of the
K$_{\alpha}$ fluorescent iron line EW ({\it left})
and the hard X-ray flux ({\it right}) between
measurements done by ASCA/BeppoSAX (y-axis)
and XMM-Newton/{\it Chandra} (x-axis). Objects
from the XMM-Newton Compton-thick sample
are indicated by {\it filled dots},
UGC~4203 (the prototypical ``Phoenix Galaxy''
with an {\it empty square}. The {\it dashed-dotted}
lines indicate a factor of $\pm3$ variation
from the equality ({\it solid}) line. \label{fig2}}
\end{figure}

\begin{figure}[ht]
\centerline{
\psfig{figure=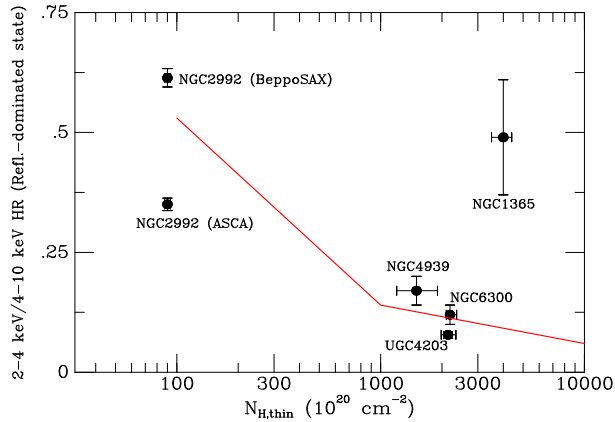,height=5.5cm,width=8.0cm,angle=-90}
}
\caption{Comparison between
the absorber column density measured during
transmission-dominated states and the
hard X-ray hardness ratio measured
during reflection-dominated states for all the
objects showing a transition. The {\it solid line}
indicates the expected relation according
to Matt et al. (2003).
Objects {\it above} the line have
never been reflection-dominated, despite
their formal classification. For objects
{\it below} the line the absorber
and the reflector cannot have the same
column density,
by contrast with a scenario whereby a
compact and homogeneous torus
covers the nucleus. To the latter
group belongs NGC~6300 as well (Guainazzi 2002),
on the basis of its $E \ge 10$~keV RXTE spectrum.
 \label{fig3}}
\end{figure}

In Fig.\ref{fig2} we show a comparison
between the ASCA/BeppoSAX and the XMM-Newton/{\it Chandra}
measurements of two of the observables, which
identify a transition:
a) a change in the Equivalent Width (EW) of
the K$_{\alpha}$ fluorescent iron line, from (less then)
$\simeq$100 to several hundreds/thousand~eV; b) hard
({\it e.g.} 2--10~keV)
X-ray flux changes by a factor $\approxgt$3--5;
The same observables are shown for
UGC~4203 as well, the prototypical ``Phoenix Galaxy''
(Guainazzi et al. 2002). Out
of 8 objects observed by XMM-Newton so far, 1
(NGC~4939) exhibits
a $\approxgt$ factor of 10
change in the K$_{\alpha}$ iron line
EW, together with a moderate variation of the
hard X-ray flux. This confirms previous results,
suggesting that these transitions
are detected in $\simeq$10\% of
nearby Seyfert~2 galaxies, when ASCA/BeppoSAX
and XMM-Newton/{\it Chandra} observations are compared.
Monte-Carlo
simulations of Compton spectra produced by slabs
with different column densities
indicate that in NGC~4939 the absorber 
in the transmission-dominated and the reflector
in the reflection-dominated state
may have the same column density (Matt et al. 2003; cf.
Fig.~\ref{fig3}).

\end{document}